\documentstyle[12pt,epsfig]{article}
\def\be{\begin{equation}} \def\ee{\end{equation}} \def\bea{\begin{eqnarray}}
\def\eea{\end{eqnarray}} \def\nnb{\nonumber}
\def\calO{{\cal{O}}}
\def\iso#1#2{\mbox{${}^{#2}{\rm #1}$}}

\def\he#1{\iso{He}{#1}}
\def\li#1{\iso{Li}{#1}}
\def\be#1{\iso{Be}{#1}}

\def\la{\mathrel{\mathpalette\fun <}}

\def\fun#1#2{\lower3.6pt\vbox{\baselineskip0pt\lineskip.9pt
  \ialign{$\mathsurround=0pt#1\hfil##\hfil$\crcr#2\crcr\sim\crcr}}}
\def\beq{\begin{equation}}
\def\eeq{\end{equation}}
\begin{document}
\hfill{TRI-PP-05-29}

\begin{center}
\vskip 4mm
{\Large\bf 
Radiative neutron capture on a proton \\
\vskip 2mm
at BBN energies
}
\vskip 1cm
{\large
S. Ando $^{a, b,}$\footnote{E-mail:sando@meson.skku.ac.kr}, 
R.~H. Cyburt$^{a,}$\footnote{E-mail:cyburt@triumf.ca}
S.~W. Hong $^{b,}$\footnote{E-mail:swhong@skku.ac.kr},
and
C.~H. Hyun $^{b, c,}$\footnote{E-mail:hch@meson.skku.ac.kr},
}\\
\vskip 7mm
{\it ${}^a$Theory Group, TRIUMF, 4004 Wesbrook Mall, Vancouver,
B.C. V6T 2A3, Canada}\\
\vskip 2mm
{\it ${}^b$Department of Physics 
and Institute of Basic Science, Sungkyunkwan University,
Suwon 440-746, Korea}\\
\vskip 2mm
{\it ${}^c$School of Physics, Seoul National University,
Seoul 151-742, Korea}\\
\end{center}

\vskip 0.5cm

The total cross section for radiative neutron capture 
on a proton, $np \to d \gamma$, 
is evaluated at big bang nucleosynthesis (BBN) energies. 
The electromagnetic transition amplitudes 
are calculated up to next-to leading order 
within the framework of pionless effective 
field theory with dibaryon fields.
We also calculate the $d\gamma\to np$ cross section
and the photon analyzing power for the $d\vec{\gamma}\to np$
process from the amplitudes.
The values of low energy constants
that appear in the amplitudes
are estimated by a Markov Chain Monte Carlo analysis 
using the relevant low energy experimental data. 
Our result agrees well with those of other theoretical calculations
except for the $np\to d\gamma$ cross section at some energies 
estimated by an R-matrix analysis.
We also study the uncertainties in our estimation
of the $np\to d\gamma$ cross section at relevant BBN energies
and find that 
the estimated cross section 
is reliable to within $\sim$1\% error.

\vskip 5mm \noindent
PACS(s): 02.50.Ga, 02.70.Uu, 11.10.Ef, 25.40.Lw, 26.35.+c

\newpage

\noindent
{\bf 1. Introduction}

Primordial nucleosynthesis happens 
between 1 and 10$^2$ seconds after
the big bang at temperatures ranging from 
$T \simeq 1$ MeV to 70 keV.  
(These are the temperatures of weak freeze
out and the end of the D bottleneck, respectively).
Predictions of primordial light element abundances, 
D, $^3$He, $^4$He and $^7$Li, and their comparison 
with observations are a crucial test of 
the standard big bang cosmology.  
The uncertainties in these predictions are dominated by the nuclear
physics input from reaction cross sections.  Reaction databases are
continuously updated~\cite{detal-XX04,setal-jcap04,c-prd04}, with more
attention now paid to the error budget.
In order to understand big bang nucleosynthesis (BBN) more clearly, it
is essential to accurately measure these reaction cross sections at
the energies relevant for BBN.  

The radiative neutron capture on a proton, $np \to d\gamma$, is one of
the key reactions for BBN, since this process is the starting point of
the synthesis of the light elements ({\it i.e.} 
it determines the end of the D bottleneck).
The cross sections of the $np \to d\gamma$ reaction have been measured
by Suzuki {\it et al.}~\cite{setal-aj95} and 
Nagai {\it et al.}~\cite{netal-prc97}.  
Its inverse process, the photo-disintegration
of the deuteron, $d\gamma \to np$, has had its cross section measured
near threshold by Hara {\it et al.}~\cite{hetal-prd03} and
Moreh {\it et al.}~\cite{moreh-prc89}\footnote{
We do not include the data from Bishop {\it et al.}~\cite{betal-pr50}
in this work because of the wrong normalization factor 
of the data~\cite{n-04}.
}, 
and the photon analyzing power for the
deuteron photo-disintegration are reported by
Schreiber {\it et al.}~\cite{setal-prc00} and 
Tornow {\it et al.}~\cite{tetal-plb03}.
Although these data comprise an important data set, 
they nevertheless only sparsely sample 
the energies relevant for BBN.
Hence an attempt at 
applying these experimental data
directly to the BBN predictions 
would make the uncertainties larger.

Theoretical calculations suggest that theory errors can be
sufficiently smaller than 
typical experimental uncertainties $\sim5\%$, 
and that they can provide a very useful discriminant for theories and
their perturbative schemes.
BBN reaction compilations adopt these theory-based cross sections
since they can provide more robust and accurate predictions than
experiment alone.
However, the uncertainties from the recent theoretical estimations 
of the cross section for $np \rightarrow d\gamma$ at BBN energies
are considerably different from each other; 4\%~\cite{cs-prc99}, 
2 $\sim$ 3\%~\cite{jh-npa01}, and 1\%~\cite{r-npa00}.
These differences could lead 
to different uncertainties in the BBN predictions,
and thus it is necessary to examine
the relevant error budget for  
the $np \rightarrow d \gamma$ process 
with a new method. 

Effective field theories (EFTs) provide 
a model-independent calculation and 
a  systematic perturbation scheme
in terms of $Q/\Lambda$ 
in calculations of various low energy hadronic 
processes~\cite{w-97,k-05},
where $Q$ is a typical momentum 
scale of a reaction in question and $\Lambda$ is a   
large scale integrated out from effective Lagrangian.
Since the energies relevant for BBN ($T\simeq 1$ MeV) are
significantly smaller than the pion mass, we can consider the pions
as heavy degrees of freedom and integrate them out of the 
Lagrangian.
Pionless EFTs~\cite{crs-npa99} 
have been intensively studied in various two-,
three- and four-nucleon processes for the last decade 
(see, {\it e.g.}, Refs.~\cite{betal-01,bk-arnps02} 
for reviews and references therein).
Convergence in the pionless EFT-based perturbative expansion
turns out to be rather slow 
for the deuteron channel 
due to a relatively large expansion parameter,
$Q/\Lambda \sim 1/3$~\cite{bbsk-npa02}. 
This large expansion parameter
essentially determines the uncertainty
estimates of the pionless EFT calculations.  
For example, in the next-to-next-to-next-to leading order (N$^3$LO) 
calculation of the $np\to d\gamma$ cross section 
at the BBN energies, Chen and Savage
found a $(1/3)^3\sim$4\% error~\cite{cs-prc99}.
Rupak pushed the calculation 
one order higher, {\it i.e.}, up to N${}^4$LO,
and found a $(1/3)^4\sim 1$\% 
theoretical uncertainty in the cross section~\cite{r-npa00}.

It has been suggested that 
the convergence of the pionless EFTs 
for the deuteron channel can be
improved by adjusting the deuteron wave function 
so as to fit it to the asymptotic
$S$-state normalization constant $Z_d=\gamma\rho_d/(1-\gamma\rho_d)$
($\gamma = \sqrt{m_N B}$ where $B$ is the binding energy of the
deuteron and $\rho_d$ is its effective range)
\cite{r-99,pc-npa00,prs-plb00}.  
By introducing dibaryon fields which represent a resonance scattering
state of the $^1S_0$ channel and the deuteron bound state of the
$^3S_1$ channel, Beane and Savage showed that the dibaryon EFT (dEFT)
without pions can naturally account for the long tail of the deuteron
wave function in the renormalized dibaryon propagator at the deuteron
pole~\cite{bs-npa01}.\footnote{ Recently, dEFT has been employed in
studying, {\it e.g.}, neutron-neutron fusion~\cite{ak-05}, muon
capture on the deuteron~\cite{yetal-05}, and the deeply-virtual
Compton scattering dissociation of the deuteron and the EMC
effect~\cite{bs-05}.}
The slow convergence problem, however, was not fully resolved in dEFT.
For instance, as discussed in detail in Ref.~\cite{ah-04}, when one
includes the electromagnetic (EM) interaction into the $np$ system in
the framework of dEFT, vector(photon)-dibaryon-dibaryon ($Vdd$)
vertices, which are classified in the subleading order in the previous
work~\cite{bs-npa01}, give contributions comparable to those of the
leading ones.
In Ref.~\cite{ah-04}, we suggested a simple prescription to extract a
LO contribution from the low energy constants (LECs) of the $Vdd$
vertices (we will mention it below), and this re-ordering of the $Vdd$
term has shown a satisfactory convergence rate similar to that
reported in other EFT calculations.
We employ this modified counting of the $Vdd$ vertex, and confirm that
this machinery is useful for the calculations of processes and
observables considered here.

In this work, we calculate the cross sections of the $np\to d\gamma$
process at BBN energies, its inverse reaction $d\gamma\to np$, and the
photon analyzing power for the $d\vec{\gamma}\to np$ process within
the framework of dEFT up to next-to leading order (NLO).
Values and uncertainties of LECs that appear in the amplitudes are
estimated by a Markov Chain Monte Carlo (MCMC) analysis using the
relevant low energy experimental data; the total cross section of the
$np$ scattering at the energies $\la$5 MeV, the rates of the $np\to
d\gamma$ process, its inverse process $d\gamma\to np$, and the photon
analyzing power in the $d\vec{\gamma}\to np$ process.
Having fitted the values of the LECs, we compare our results with the
experimental data mentioned above, and find that our statistical error
bars of the $np\to d\gamma$ cross section 
are satisfactorily improved compared to the experimental ones.
We also compare our result of the $np\to d\gamma$ cross section with
other theoretical estimations, the pionless EFT calculation up to
N$^4$LO~\cite{r-npa00}, a calculation with Argonne v18 (Av18)
potential and the meson exchange current~\cite{nakamura04}, and the
result of an R-matrix analysis~\cite{hale04}.
Our result agrees quite well with those of the previous EFT and Av18
potential model calculations within the uncertainties ($\sim$1\%)
estimated by MCMC.  
On the other hand we find significant ($\sim$4.6\%) difference
in the $np\to d\gamma$ cross sections from the R-matrix theory
estimated at $E=$ 0.1 and 1 MeV where $E$ is the total two-nucleon
kinetic energy in the center of mass frame.
Since the various theoretical uncertainties from higher order
corrections in the pionless EFT calculation have already been studied
in Ref.~\cite{r-npa00} and the slow convergence problem in the former
pionless EFT calculations has been resolved in dEFT, we discuss that
the theoretical estimations of the $np\to d\gamma$ cross section at
the BBN energies are reliable with an uncertainty of $\la$1\%.

This paper is organized as follows. In Sect.~2, we present the
pionless effective Lagrangian with dibaryon fields up to NLO.  We
calculate the amplitudes with the $S$- and $P$-wave $np$ states and
the cross sections for the $np \to d\gamma$ and $d\gamma \to np$
processes and the photon analyzing power for the $d\vec{\gamma}\to np$
process up to NLO in Sect.~3.
Utilizing a Markov Chain algorithm, we
determine the values and uncertainties of the 
LECs in Sect.~4.   
In Sect.~5, we compare the resulting 
observables to the experimental data 
and other theoretical estimations.
In Sect.~6,  
we summarize our results and give discussion.
In appendix A, 
we present the expressions of 
the renormalized dibaryon propagator 
and the $S$-wave $NN$ scattering amplitudes,
and show the relations between the LECs in the strong sector
and the parameters in the effective range theory.
In appendix B, 
we describe in detail the MCMC analysis in 
determining the LECs and cross sections.

\vskip 3mm \noindent
{\bf 2. Pionless effective Lagrangian with dibaryon fields}

A pionless effective Lagrangian 
for nucleon and dibaryon fields
interacting with an external vector field 
can be written as~\cite{bs-npa01,ah-04}
\bea
{\cal L} = {\cal L}_N
+ {\cal L}_s
+ {\cal L}_t
+ {\cal L}_{st} \, ,
\eea
where ${\cal L}_N$ is the nucleon Lagrangian, ${\cal L}_s$ and ${\cal
L}_t$ are the Lagrangian for the dibaryon fields in ${}^1S_0$ and
${}^3S_1$ channels, respectively.
${\cal L}_{st}$ is the Lagrangian that accounts for the isovector EM
interaction of the dibaryon fields.

${\cal L}_N$ in the heavy-baryon formalism reads
\bea
{\cal L}_N =
N^\dagger \left\{
iv\cdot D
+\frac{1}{2m_N}\left[
(v\cdot D)^2-D^2  
-i[S^\mu,S^\nu]
\left((1+\kappa_V)f^+_{\mu\nu}
+(1+\kappa_S)f^S_{\mu\nu}\right)
\right] \right\} N\, ,
\label{eq;L1}
\eea
where $v^\mu$ is the velocity vector satisfying $v^2=1$;
we take $v^\mu=(1,\vec{0})$.
$S^\mu$ is the spin operator $2S^\mu=(0,\vec{\sigma})$.
$D_\mu=\partial_\mu -\frac{i}{2}\vec{\tau}\cdot\vec{\cal V}_\mu
-\frac{i}{2}{\cal V}^S_\mu$, where
$\vec{\cal V}_\mu$ and ${\cal V}_\mu^S$ are
the external isovector and isoscalar
vector currents, respectively.
$f^+_{\mu\nu} = \frac{\vec{\tau}}{2} \cdot (
\partial_\mu \vec{\cal V}_\nu -\partial_\nu\vec{\cal V}_\mu )$ and
$f^S_{\mu\nu} = \frac12(
\partial_\mu {\cal V}^S_\nu - \partial_\nu {\cal V}^S_\mu )$.
$m_N$ is the nucleon mass
and $\kappa_V$ ($\kappa_S$) is
the isovector (isoscalar)
anomalous magnetic moment of the nucleon,
$\kappa_V= 3.706$ ($\kappa_S= -0.120$).

${\cal L}_{s}$, ${\cal L}_t$,
and ${\cal L}_{st}$
for the dibaryon and two nucleon fields read
\bea
{\cal L}_s &=&
\sigma_s s_a^\dagger\left[iv\cdot D
+\frac{1}{4m_N}[(v\cdot D)^2-D^2]
+\Delta_s\right] s_a 
-y_s\left[s_a^\dagger (N^TP^{(^1S_0)}_aN)
+ \mbox{h.c.}\right],
\label{eq;Ls}
\\
{\cal L}_t &=&
\sigma_t t_i^\dagger\left[iv\cdot D
+\frac{1}{4m_N}[(v\cdot D)^2-D^2]
+\Delta_t
\right] t_i
-y_t\left[t_i^\dagger (N^TP_i^{(^3S_1)}N) + \mbox{h.c.}\right]
\nnb \\ &&
+\left[
\frac{1+\kappa_S}{2m_N}
- \frac{2l_2}{m_N\rho_d}
\right]
 i\epsilon_{ijk}t_i^\dagger t_j B_k \, ,
\label{eq;Lt}
\\
{\cal L}_{st} &=&
\left[
-\frac{1+\kappa_V}{2m_N}
\left(\frac{r_0+\rho_d}{2\sqrt{r_0\rho_d}}\right)
+\frac{l_1}{m_N\sqrt{r_0\rho_d}}
\right]
\left( t_i^\dagger s_3 B_i
+ \mbox{h.c.}\right) \, .
\label{eq;Lst}
\eea
The covariant derivative for the dibaryon field 
is given by
$D_\mu =\partial_\mu -iC{\cal V}^{ext}_\mu$, where
${\cal V}_\mu^{ext}$ is the external vector field
and $C$ is the charge operator of the dibaryon fields;
$C=0,1,2$ for $nn$, $np$, $pp$ channels, respectively,
and we have set $e=1$.
$\vec{B}$ is the magnetic field given by
$\vec{B}=\vec{\nabla} \times\vec{\cal V}^{ext}$.
$\sigma_{t}$ ($\sigma_s$) is the sign factor,
$\Delta_t$ ($\Delta_s$) is the difference
between the dibaryon mass $m_t$ ($m_s$)
in the $^3S_1$ ($^1S_0$) channel
and the two-nucleon mass;
$m_{t,s} = 2 m_N + \Delta_{t,s}$,
and $y_{s,t}$ are the dibaryon-nucleon-nucleon ($dNN$) 
coupling constants of 
the dibaryon spin singlet and triplet channels.
In appendix A, these LECs in the strong sector 
are related to 
the parameters of the effective range theory.
$l_1$ and $l_2$ are LECs of the $Vdd$ vertices and 
can be fixed, for instance, by the thermal $np\to d\gamma$ rate and 
the deuteron magnetic moment, respectively.  
We note that we have separated the leading contributions from
the coefficients in the $Vdd$ vertices associated with $l_1$ and $l_2$
and fixed them mainly by one-body interactions, {\it i.e.}, the
vector(photon)-nucleon-nucleon ($VNN$) couplings, as suggested in
Ref.~\cite{ah-04}.
Consequently, $l_1$ and $l_2$ terms 
in this work give genuine NLO contributions.
$\rho_d$ and $r_0$ are
the effective ranges for the deuteron
and $^1 S_0$ scattering state, respectively.
$P_i^{({\cal S})}$ is the projection operator
for the ${\cal S} ={}^{2S+1}L_J$ channel.
For the $S$ and $P$ waves which are dominant
at low energies, 
the projection operators are given as~\cite{fms-npa00};
\bea
&&
P_i^{({}^3S_1)} = \frac{1}{\sqrt{8}}\sigma_2\sigma_i\tau_2,
\ \ \ 
P_a^{({}^1S_0)} = \frac{1}{\sqrt{8}}\sigma_2\tau_2\tau_a,
\ \ \ 
P^{(^1P_1)}_a = \sqrt{\frac38} \hat{p}\cdot \vec{\epsilon}
\tau_a \, ,
\nnb \\ &&
P^{(^3P_0)}_a = 
\frac{1}{\sqrt{8}}
\sigma_2 \vec{\sigma}\cdot \hat{p}
\tau_2\tau_a\, , 
\ \ \ 
P^{(^3P_1)}_a =
\frac{\sqrt{3}}{4}
\epsilon^{ijk}\epsilon^i \hat{p}^j\sigma_2\sigma^k
\tau_2 \tau_a \, ,
\nnb \\ &&
P^{(^3P_2)}_a = 
\sqrt{\frac38}\epsilon^{ij}\hat{p}^i
\sigma_2\sigma^j
\tau_2\tau_a\, ,
\ \ \ 
\int \frac{d\Omega_{\hat{p}}}{4\pi}
{\rm Tr}(P_i^{({\cal S})\dagger} P^{({\cal S})}_j) = \frac12\delta_{ij},
\eea
where $\epsilon^i$ and $\epsilon^{ij}$ are 
$J$=1 and 2 polarization tensor, respectively,
and
$\sigma_i$ ($\tau_a$) with $i(a)=1,2,3$ 
is the spin (isospin) operator.

We adopt the standard counting rules of dEFT~\cite{bs-npa01}.
Introducing an expansion scale $Q<\Lambda$ $(\simeq m_\pi)$, we count
magnitude of spatial part of the external and loop momenta,
$|\vec{p}|$ and $|\vec{l}|$, as $Q$, and the time component of them,
$p^0$ and $l^0$, as $Q^2$.  Thus the nucleon and dibaryon propagators
are of $Q^{-2}$ and a loop gives a factor of $Q^5$ due to the
4-dimensional differential volume in the loop integration.  The
scattering lengths and effective ranges are counted as $Q\sim
\{\gamma,1/a_0,1/\rho_d,1/r_0\}$. 
This ensures, as discussed in the introduction, that 
one reproduces the long tail of the deuteron wavefunction
characterized by $\sqrt{Z_d}$ and thus has good  
convergence~\cite{r-99,pc-npa00,prs-plb00}.
Orders of vertices and diagrams are
easily obtained by counting the numbers of these factors.

\vskip 3mm \noindent
{\bf 3. Amplitudes and cross sections}

\begin{figure}[tbp]
\begin{center}
\epsfig{file=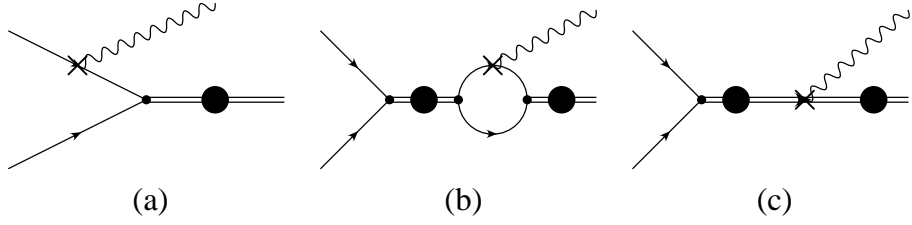,width=12cm}
\caption{\label{fig;npdg}
Diagrams contributing to $np\to d\gamma$
and $d\gamma\to np$:
Diagrams (a) and (b) are of LO, $\calO(Q^{1/2})$,
while diagram (c) consists of LO and NLO terms, $\calO(Q^{3/2})$.
The single lines,
the double lines with a filled 
circle (see Fig.~\ref{fig;dprop} as well), 
and the wavy lines
denote nucleons, renormalized dibaryons, and photons, respectively.
$VNN$ vertex ``$\times$" in (a) and (b) 
and the LO part of $Vdd$ vertex ``$\times$'' in (c)
are proportional 
to ($1+\kappa_S$) and ($1+\kappa_V$) for the initial 
$^3S_1$ and $^1S_0$ channel, respectively.
LECs $l_1$ and $l_2$ appear in 
the NLO part of the $Vdd$ vertex ``$\times$'' 
in the diagram (c).
}
\end{center}
\end{figure}

Diagrams for $np \rightarrow d\gamma$ up to NLO 
are depicted in Fig.~\ref{fig;npdg}.
Diagrams (a) and (b) give only LO contributions. 
Diagram (c) consists of LO and NLO fractions
and the NLO contributions from the diagram (c) stem from 
the LECs $l_1$ and $l_2$, 
as discussed above.

Summing up the contributions 
of the diagrams (a), (b) and (c)
in Fig. \ref{fig;npdg},
we obtain the amplitudes
for the initial $^3S_1$ and $^1S_0$ states as~\cite{bs-npa01,ah-04}
\bea
\lefteqn{
i{\cal A}^{(a+b+c)}({}^3S_1) =
-i(\vec{\epsilon}_{(d)}^*\times\vec{\epsilon}_i)
\cdot (\vec{\epsilon}_{(\gamma)}^*\times \hat{k})
} \nnb \\ && \times 
\sqrt{\frac{2\pi\gamma}{1-\gamma\rho_d}}
\frac{2}{m_N}
\frac{1}{-\gamma-ip+\frac12 \rho_d(\gamma^2+p^2)}
\frac{\gamma^2+p^2}{m_N}l_2 \, , 
\label{eq;A3S1}
\\
\lefteqn{
i{\cal A}^{(a+b+c)}({}^1S_0) =
\vec{\epsilon}_{(d)}^*\cdot
(\hat{k}\times\vec{\epsilon}_{(\gamma)}^*)
\sqrt{\frac{2\pi\gamma}{1-\gamma\rho_d}}
\frac{2}{m_N}
\frac{1}{-\frac{1}{a_0}-ip+\frac12 r_0p^2}
} \nnb \\ && \times
\left\{
\frac{1+\kappa_V}{2m_N}\left[
\gamma-\frac{1}{a_0}
-\frac14 (r_0+\rho_d) \gamma^2
+\frac14 (r_0-\rho_d) p^2
\right]
+ \frac{\gamma^2+p^2}{2m_N} l_1
\right\}\, .
\label{eq;A1S0}
\eea
Here $2\vec{p}$ is the relative momentum of the two-nucleon
system, and $\vec{k}$ is the momentum of the outgoing photon;
$p=|\vec{p}|$,
$k=|\vec{k}|$,
and $\hat{k}=\vec{k}/k$.
$\vec{\epsilon}_{(d)}^*$ and $\vec{\epsilon}_{(\gamma)}^*$
are the polarization vectors for the outgoing deuteron
and photon, respectively. 
$a_0$ is the scattering length in the ${}^1S_0$ channel.
The LECs in the strong sector have been renormalized
by the effective range parameters.
See appendix A for details.
We note that the isovector $M1$ amplitude
(the ${}^1S_0$ channel), Eq.~(\ref{eq;A1S0}),
has contributions from both LO and NLO.
Whereas, the isoscalar $M1$ amplitude
(the ${}^3S_1$ channel), 
Eq.~(\ref{eq;A3S1}), has no LO contribution
due to the orthogonality between the bound and 
scattering states for the ${}^3S_1$ channel, 
but non-zero amplitude 
proportional to $l_2$
appears at NLO.

From the diagram (a) in Fig.~\ref{fig;npdg},
amplitudes with initial $P$-waves are obtained as;
\bea
i{\cal A}^{(a)}({}^3P_0) &=& 
-i\vec{\epsilon}^*_{(\gamma)}\cdot \vec{\epsilon}^*_{(d)}
\sqrt{\frac{2\pi \gamma}{1-\gamma\rho_d}}
\frac23\frac{p}{m_N(\gamma^2+p^2)}\, ,
\label{eq;A3P0}
\\
i{\cal A}^{(a)}({}^3P_1) &=& 
-i\vec{\epsilon}_1\cdot(\vec{\epsilon}^*_{(\gamma)}\times\vec{\epsilon}^*_{(d)})
\sqrt{\frac{2\pi \gamma}{1-\gamma\rho_d}}
\sqrt{\frac23}\frac{p}{m_N(\gamma^2+p^2)}\, ,
\label{eq;A3P1}
\\
i{\cal A}^{(a)}({}^3P_2) &=& 
\epsilon_2^{ij}\epsilon^{*i}_{(\gamma)}\epsilon^{*j}_{(d)}
\sqrt{\frac{2\pi \gamma}{1-\gamma\rho_d}}
\frac{2}{\sqrt3}\frac{p}{m_N(\gamma^2+p^2)}\, .
\label{eq;A3P2}
\eea 
We note that
the diagrams (b) and (c) in Fig.~\ref{fig;npdg} do not contribute 
to the $P$-wave amplitudes 
because the states of 
the renormalized dibaryon propagator 
are only the $S$-waves.

Having the amplitudes calculated above,
we can easily obtain 
the expressions of the cross sections
of the $np\to d\gamma$ and $d\gamma\to np$
processes and the analyzing power for the $d\vec{\gamma}\to np$
process. 
The total cross section of the $np\to d\gamma$
process in the CM frame reads~\cite{bs-npa01,ah-04}
\bea
\sigma =
\frac{\alpha(\gamma^2+p^2)}{4 p}
\sum_{\rm spin}|{\cal A}|^2 ,
\label{eq;sig-npdg}
\eea
where $\alpha$ is the fine structure constant.
\footnote{
In obtaining the total cross section,  
the following identities are useful.
\bea
\sum_{\rm spin}|i\vec{\epsilon}_{(d)}^*\cdot
(\hat{k}\times\vec{\epsilon}_{(\gamma)}^*)|^2 = 2\, ,
\ \ 
\sum_{\rm spin}|\vec{\epsilon}^*_{(\gamma)}\cdot \vec{\epsilon}^*_{(d)}|^2=2,
\ \ 
\sum_{\rm spin}|\vec{\epsilon}_1\cdot 
(\vec{\epsilon}^*_{(\gamma)}\times\vec{\epsilon}^*_{(d)})|^2 = 4,
\ \ 
\sum_{\rm spin}|\epsilon_2^{ij}\epsilon^{*i}_{(\gamma)}\epsilon^{*j}_{(d)}|^2
= \frac{10}{3}\, .
\nnb 
\eea
We ignore the amplitude for the $^3S_1$ channel in
Eq.~(\ref{eq;A3S1}), as will be discussed later.}

The total cross section of the photo-disintegration 
of the deuteron, $d\gamma\to np$,
has a simple relation with the cross section of 
its inverse process as~\cite{cs-prc99}
\bea
\sigma(\gamma d\to np) 
= \frac{2m_N(E_\gamma-B)}{3E_\gamma^2}\sigma(np\to d\gamma)\, ,
\label{eq;sig-dgnp}
\eea
where $E_\gamma$ is the photon energy.
Since we have already obtained 
the $np$-capture cross section, 
the calculation of $\sigma(\gamma d\to np)$
is straightforward. 

The photon analyzing power $\Sigma(\theta)$
with the linearly polarized photons
in the $d\vec{\gamma}\to np$ process
is defined as
$\Sigma(\theta) \equiv (N_\parallel-N_\perp)/(N_\parallel+N_\perp)$
where $N_\parallel$ and $N_\perp$ are the number of outgoing 
neutrons counted 
in and out of the horizontal $\gamma$-ray 
polarization plane, respectively,
and $\theta$ is the angle between 
the incoming photon and outgoing neutron 
in the laboratory frame.
This quantity is related to 
the $M1$ and $E1$ contributions to
the total cross sections
of the $d\gamma\to np$ process,
$\sigma_{M1}$ and $\sigma_{E1}$,
respectively. 
The relation can be found 
{\it e.g.}, in Eq.~(4) of
Ref.~\cite{setal-prc00}, which reads
\bea
\Sigma(\theta) &=& \frac{\frac32 \sigma_{E1}{\rm sin}^2\theta}{
\sigma_{M1}+\frac32\sigma_{E1}{\rm sin}^2\theta} \, .
\eea
$\sigma_{M1}$ and $\sigma_{E1}$
are easily calculated 
by using Eqs.~(\ref{eq;sig-npdg}, \ref{eq;sig-dgnp}) 
and the expressions of the $M1$ and $E1$ amplitudes
obtained in Eqs.~(\ref{eq;A3S1}, \ref{eq;A1S0},
\ref{eq;A3P0}, \ref{eq;A3P1}, \ref{eq;A3P2}).

\vskip 3mm \noindent
{\bf 4. Parameter determination from experimental data}

In this section, 
we determine the values and uncertainties of the LECs 
that appear in our results from 
the relevant low energy experimental data.  
The cross sections and photon analyzing power obtained
from the amplitudes in Eqs.~(\ref{eq;A3S1}, \ref{eq;A1S0},
\ref{eq;A3P0}, \ref{eq;A3P1}, \ref{eq;A3P2}) depend
on six parameters; $a_0$, $r_0$, $\gamma$, $\rho_d$, 
$l_1$, and $l_2$.
$\gamma$ and $l_2$ are precisely determined by the
deuteron binding energy $B$ and magnetic moment $\mu_d$, 
respectively.  
Using the relation,
$\mu_d=1+\kappa_S +Z_d l_2$,
we have $l_2=-0.0154$ fm. 
It is $\sim$50 times smaller than 
the value of $l_1$ in Table~\ref{tab:parameters}
(and the expressions for the $l_1$ and $l_2$ terms 
in the amplitudes are almost identical).
We will consider the error of the cross section
at the order of 0.1\%,
so we neglect the amplitude in Eq.~(\ref{eq;A3S1})
from the ${}^3S_1$ channel in the following calculations.
Thus, the parameters $a_0$, $r_0$, $\rho_d$, 
and $l_1$ are determined 
from the low energy experimental data, 
whereas $\gamma$ is fixed from $B$ and treated as a constraint
in the fitting.

We fit these parameters to the low energy $np$ data
with a Markov Chain Monte Carlo (MCMC) analysis
using the Metropolis algorithm~\cite{metrop}.  
This method is 
more efficient in exploring parameter space than creating a
multi-dimensional grid of parameter values and interpolating 
to find the underlying likelihood distribution and 
favored parameter values.
The method is a random walk constrained to ``walk'' in regions 
of low $\chi^2$. As the $\chi^2$ increases for a particular step, 
the probability of the step being accepted into the chain decreases.  
This ``walk'' explores the allowed parameter space, and if long enough
will explore all channels of parameter degeneracy.  
The resulting sample, then provides a direct probe of the parameter
likelihood, and can be used to determine, {\it e.g.},  
the means, standard
deviations and correlations of the parameters.  
It also provides an accurate way of propagating the uncertainties and
correlations of the parameters into the cross section uncertainties,
which depend non-linearly on the parameters.  
One drawback is that this method is not particularly efficient at
sampling multi-peaked distributions.  The random walk can become
trapped in a local minimum.  For the case we are considering, the
5-dimensional likelihood distribution 
is singly peaked ({\it i.e.} a unique solution exists).

We now list the low energy experimental data employed in our fitting.
To include the high precision measurements of the deuteron binding
energy, we adopt the value of
$B=2.2245671\pm0.0000042$ 
MeV\footnote{The quoted deuteron binding energy is a
weighted average of the available measurements.  The error is the
weighted dispersion about the mean, recommended by~\cite{c-prd04,CFO1}
because standard techniques underestimate uncertainties when data 
are discrepant.}~\cite{dbind} 
as the additional constraint mentioned above.  
This accurate value of $B$ gives the uncertainty of $\gamma$
in the order of 10$^{-6}$,
thus it is effectively fixed, 
independent from the other constraints.
The $np$-scattering data are found on the NN-Online website~\cite{nn-online}. 
We restrict ourselves to the
angle-integrated scattering cross sections with center-of-mass
energies $\la$5 MeV, and have
$N_{sct}= 2124$ data points. 
We adopt the 2 thermal neutron
capture cross sections from Cox {\it et al.}~\cite{cetal-np65} 
(334.2$\pm$0.5 mb) and Cokinos and Melkonian~\cite{cm-prc77} 
(332.6 $\pm$ 0.7 mb),\footnote{
In a private
communication mentioned in~\cite{callerame}, the Cox {\it et al.}'s 
thermal cross section is renormalized from 334.2 to 333.9 mb.  
This likely indicates hidden systematics in the experiment and 
subsequently larger (and unacknowledged) uncertainties, 
though we adopt the original published number.
In addition, the thermal $np$ capture rate,
$\sigma= (334\pm 3)$ mb, 
has been estimated by T.-S. Park {\it et al.}
in the calculation of pionful 
effective field theory~\cite{tmr-prl95}.
}
and also adopt the $np$-capture data of
Suzuki {\it et al.}~\cite{setal-aj95} 
and Nagai {\it et al.}~\cite{netal-prc97}, 
the photo-dissociation cross section data 
by Hara {\it et al.}~\cite{hetal-prd03} and
Moreh {\it et al.}~\cite{moreh-prc89}, and 
the photon analyzing power data from
Schreiber {\em et al}.~\cite{setal-prc00} 
and Tornow {\em et al}.~\cite{tetal-plb03}.

\begin{figure}[h]
\epsfig{file=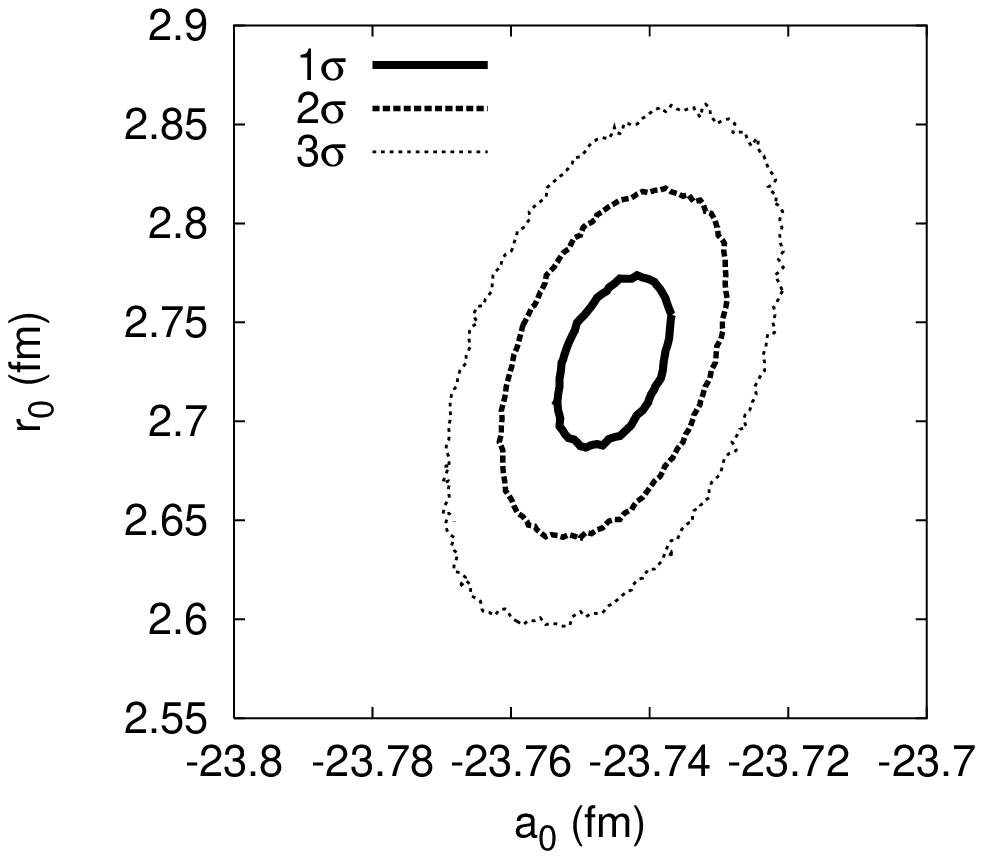,width=5.2cm}
\epsfig{file=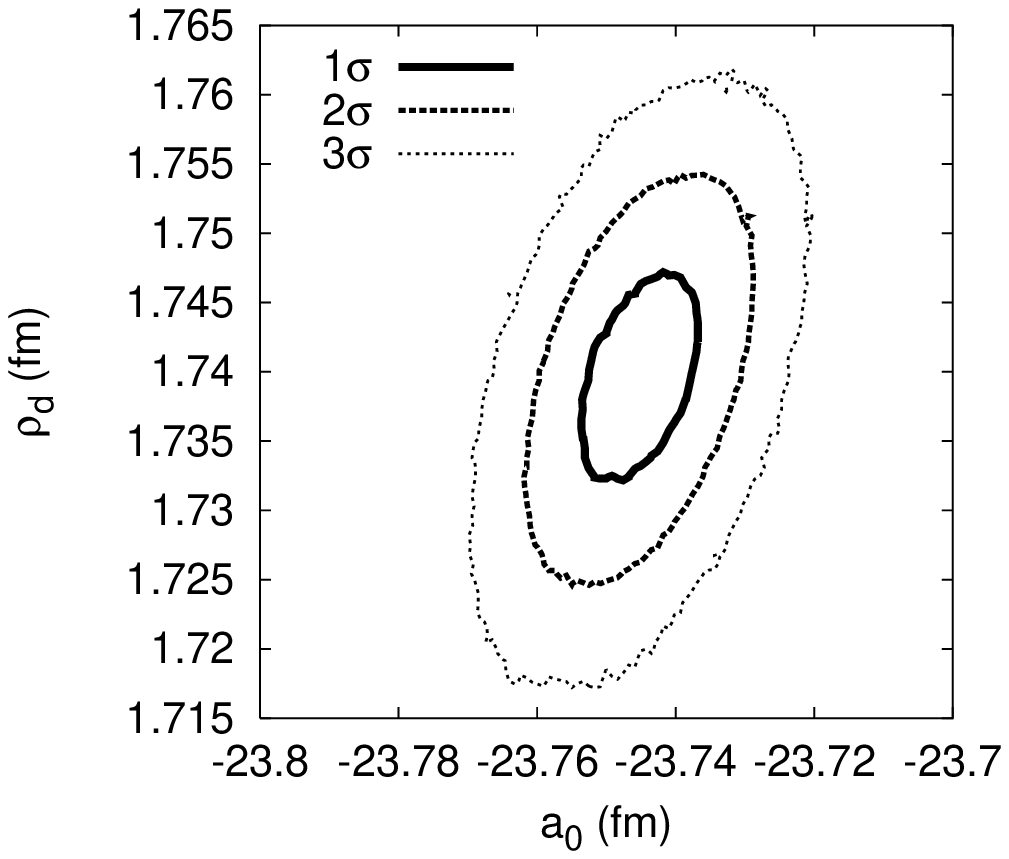,width=5.2cm}
\epsfig{file=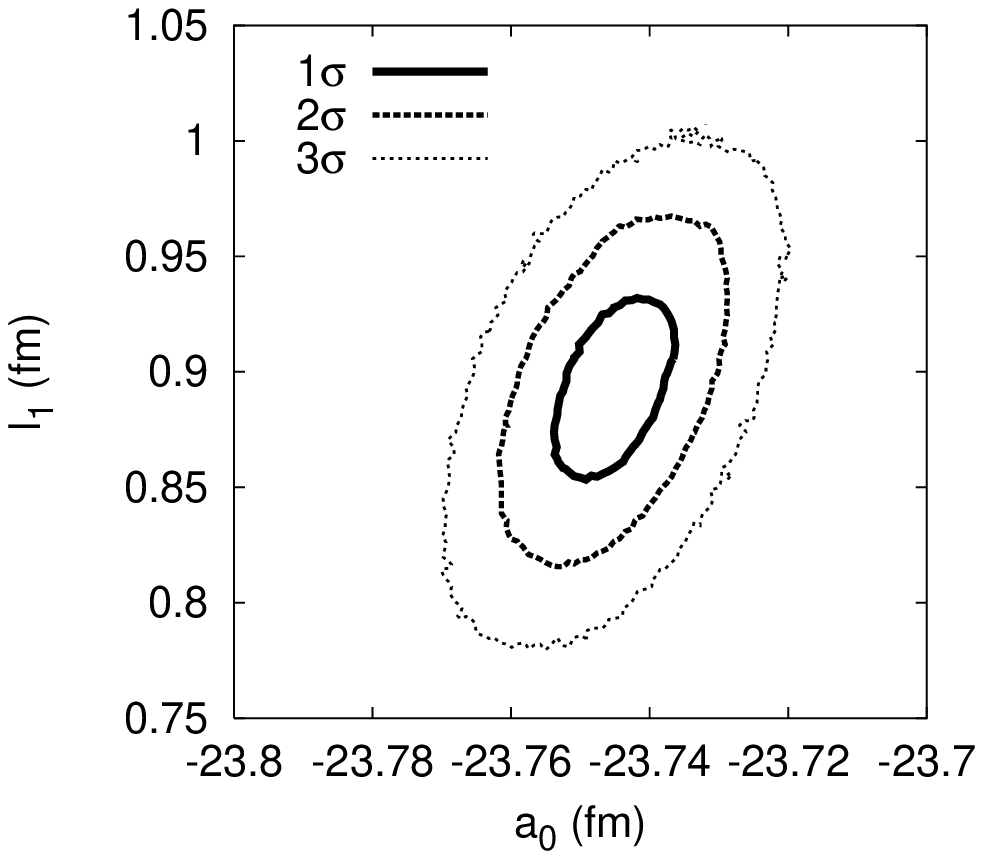,width=5.2cm}

\epsfig{file=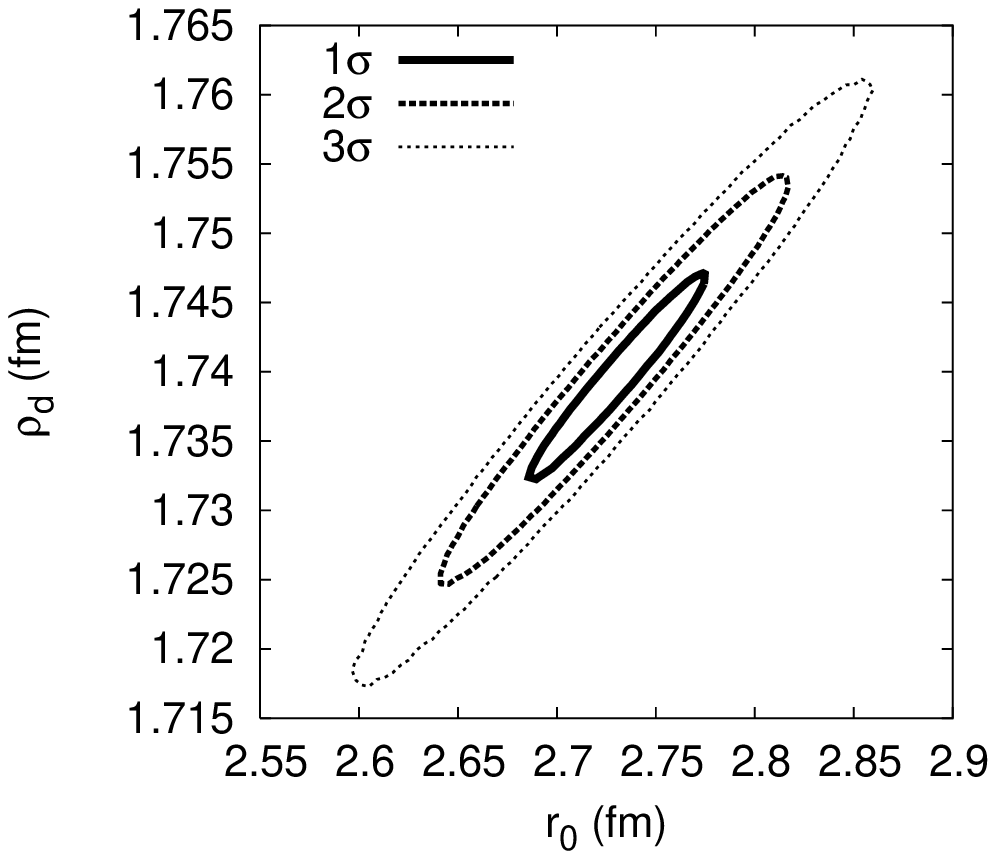,width=5.2cm}
\epsfig{file=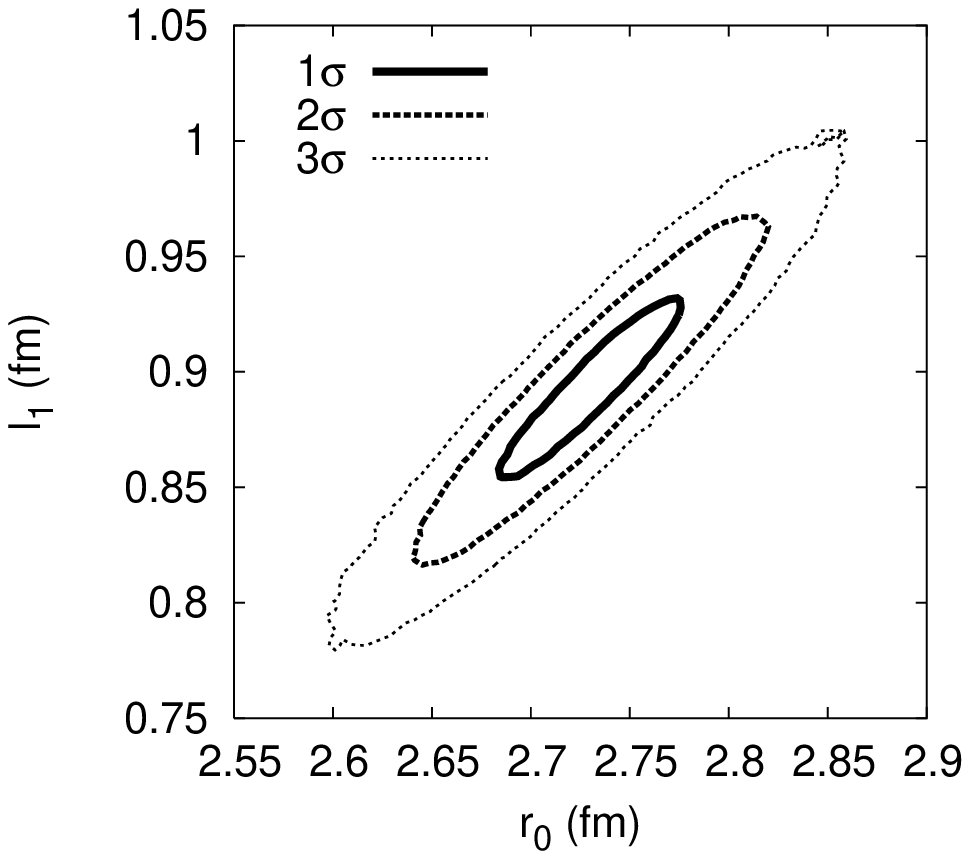,width=5.2cm}
\epsfig{file=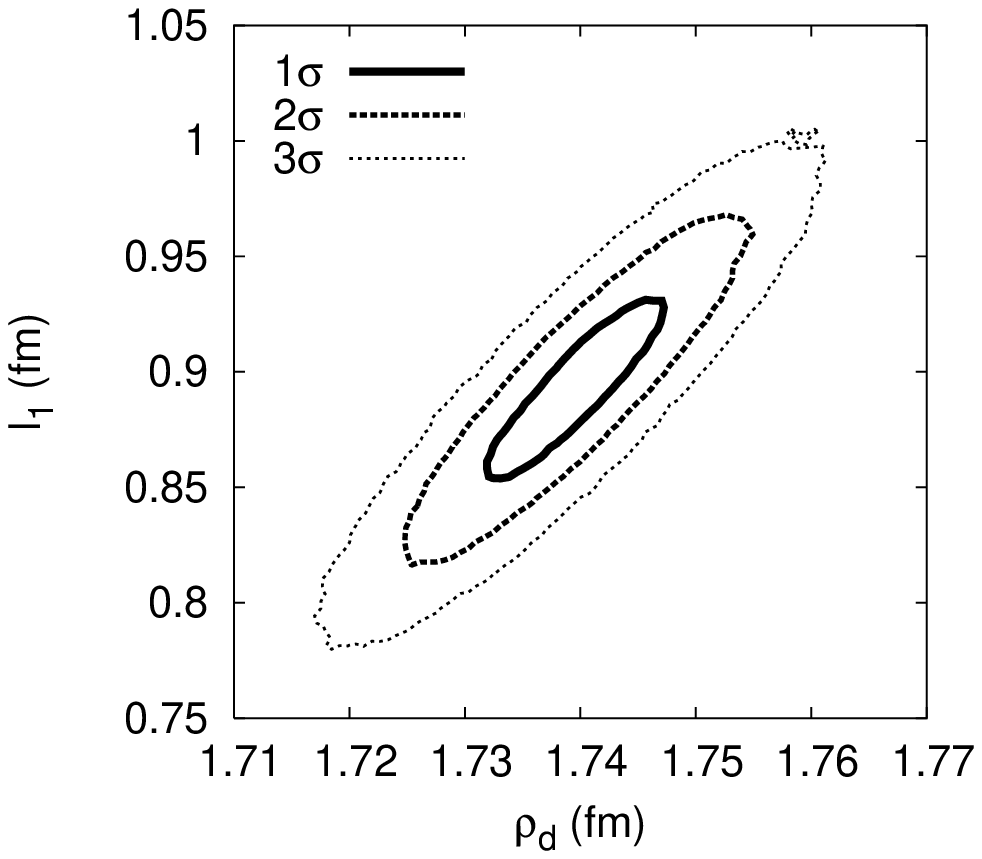,width=5.2cm}
\caption{
Contour plots of two dimensional parameter spaces 
for 5 dimensional probability distribution
generated by the MCMC analysis.  
We draw the contours of 1,2, and 3 $\sigma$ for each set of the 
parameters, $a_0, r_0, \rho_d,$, and $l_1$, whereas $B$ 
is effectively fixed at the observed value and 
uncorrelated with the other parameters, and thus not shown.
}
\label{fig;contour}
\end{figure}
\begin{table}
\begin{center}
\begin{tabular}{lrr|cccc}\hline
 & & &\multicolumn{4}{|c}{Correlations} \\ 
 & \multicolumn{1}{r}{Prev. Meth.} & \multicolumn{1}{r|}{MCMC}& 
$a_0$ &  $r_0$ &$\rho_d$& $l_1$  \\ \hline
$a_0$ & $-23.749\pm0.008$ & $-23.745\pm$0.008 & 1.000 & 0.433 & 0.464 & 0.496 \\
$r_0$ & $2.81\pm0.05$ & 2.730$\pm$0.044 & 0.433 & 1.000 & 0.975 & 0.936 \\
$\rho_d$ & $1.760\pm0.005$ & 1.740$\pm$0.007 & 0.464 & 0.975 & 1.000 & 0.898 \\
$l_1$ & $0.782\pm0.022$ & 0.893$\pm$0.038 & 0.496 & 0.935 & 0.898 & 1.000 \\
\hline
\end{tabular}
\caption{
Values and correlations of the parameters.
The values of the parameters obtained are
in units of fm.  
``Prev. Meth.'' are the adopted parameter values
from previous works; of these $a_0$, $r_0$, and $\rho_d$ are from
Ref.~\cite{nuclth-STS} and $l_1$ from the thermal $np$ capture
rates~\cite{ah-04}.  
``MCMC'' is obtained by the MCMC analysis ($\chi^2$-fit) by using the
low energy experimental data. Also shown are the parameter
correlations.  They are defined by $COR(i,j) =
COV(i,j)/(\sigma_{\rm stat.}(i)\sigma_{\rm stat.}(j))$, 
where $COV(i,j)$ is the covariance
between the $i^{\rm th}$ and $j^{\rm th}$ parameters and 
$\sigma_{\rm stat.}(i)$ is the statistical uncertainty of the
$i^{\rm th}$ parameter.}
\label{tab:parameters}
\end{center}
\end{table}
Using the experimental data mentioned above, we fit these parameters
employing the MCMC analysis.  The steps in doing the MCMC analysis are
described in detail in appendix B.
We initialize the Markov chain at the point in 5 dimensional 
parameter space that minimizes the $\chi^2$.
After verifying that the $\chi^2$ had only one minimum, 
we did not need to run more than one Markov chain, 
making the analysis much more efficient.
We display, 
in Fig.~\ref{fig;contour},
two dimensional contour plots 
of 1, 2, and 3 $\sigma$ for each set of parameters, 
$a_0$, $r_0$, $\rho_d$, $l_1$ 
from our 5 dimensional probability
distribution generated from the MCMC analysis.   
We note that since $B$ is effectively fixed and 
uncorrelated with the other parameters, 
contour plots with $B$ are not necessary.
``MCMC'' in Table~\ref{tab:parameters} 
are the average parameter values and standard deviations from this
analysis with the total number of the data $N_{\rm tot}=2147$ and the
minimum $\chi^2$, $\chi^2_{\rm tot,min} =2303.00$. (For more details, see
Table~\ref{tab:minchi2} in the appendix B.)
These parameter values agree well with prior determinations shown in
``Prev. Meth.'' in the same table.
The small ($\la$ 2\%) differences between 
``MCMC" and ``Prev. Meth."
could be interpreted as effects from either higher order corrections not
included in our calculation or simply statistical fluctuations.
Also in Table~\ref{tab:parameters},
we show the correlations of the parameters.
We find weak correlations of $a_0$ with the three parameters
$r_0$, $\rho_d$, and $l_1$ and strong correlations among these three
parameters.
Though we formally counted the four parameters as
in the same order in this work, 
this may indicate the existence of a perturbative series 
that 
the contribution of $a_0$ is the LO one
and the three parameters are in the same order (NLO), 
as already discussed in Ref.~\cite{bbsk-npa02}. 

\vskip 3mm \noindent
{\bf 5. Numerical results}

\begin{figure}
\begin{center}
\epsfig{file=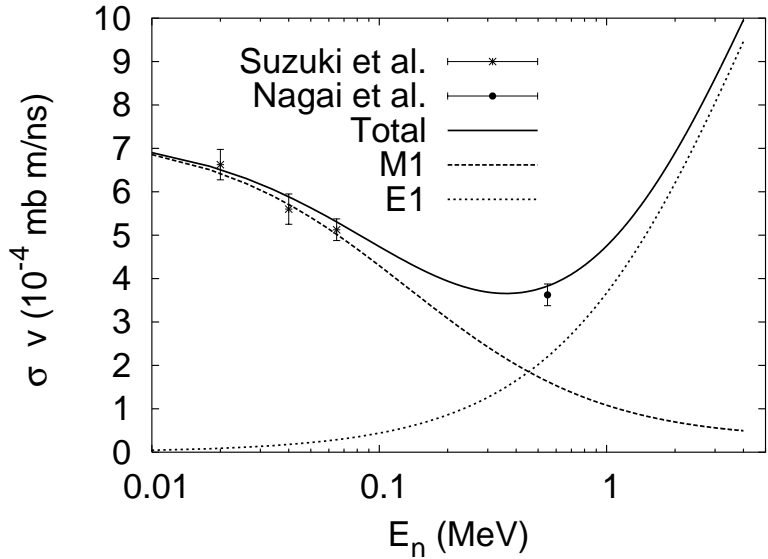, width=11cm}
\end{center}
\caption{
Total cross section of $n + p \rightarrow d + \gamma$ in units of mb
multiplied by neutron speed in m/ns. Neutron energy $E_n$ is in
the laboratory frame. Dashed and dotted curves are the $M1$ and $E1$
contributions to the total cross section, respectively.  The
experimental data are from Suzuki {\it et al.}~\cite{setal-aj95} and
Nagai {\it et al.}~\cite{netal-prc97}.}
\label{fig;CSnpdg}
\end{figure}

\begin{figure}
\begin{center}
\epsfig{file=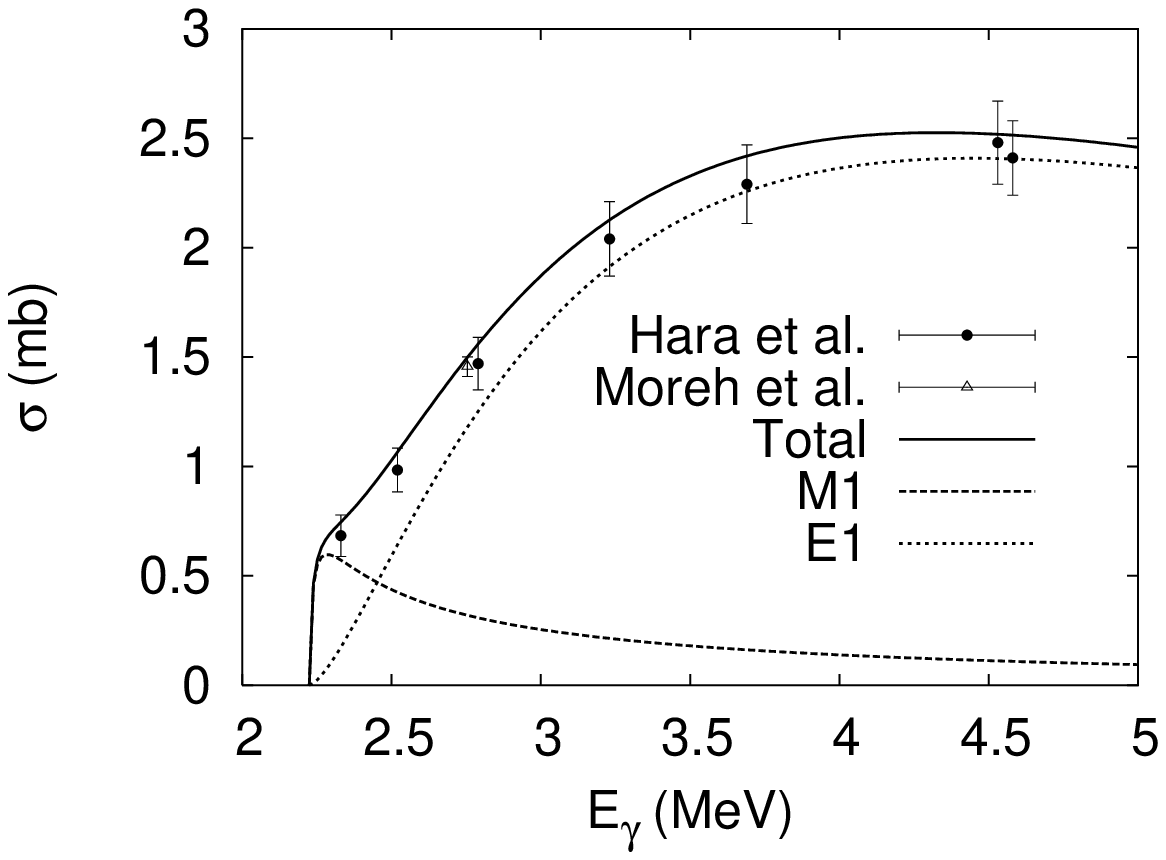, width=11cm}
\end{center}
\caption{
Total cross section for the $d + \gamma \rightarrow n + p$ process.
Dashed and dotted curves represent the $M1$ and $E1$ contributions to the
total cross section, respectively.  The experimental data are from
Hara {\it et al.}~\cite{hetal-prd03} and Moreh {\it et
al.}~\cite{moreh-prc89}.}
\label{fig;dgnp}
\end{figure}

\begin{figure}
\begin{center}
\epsfig{file=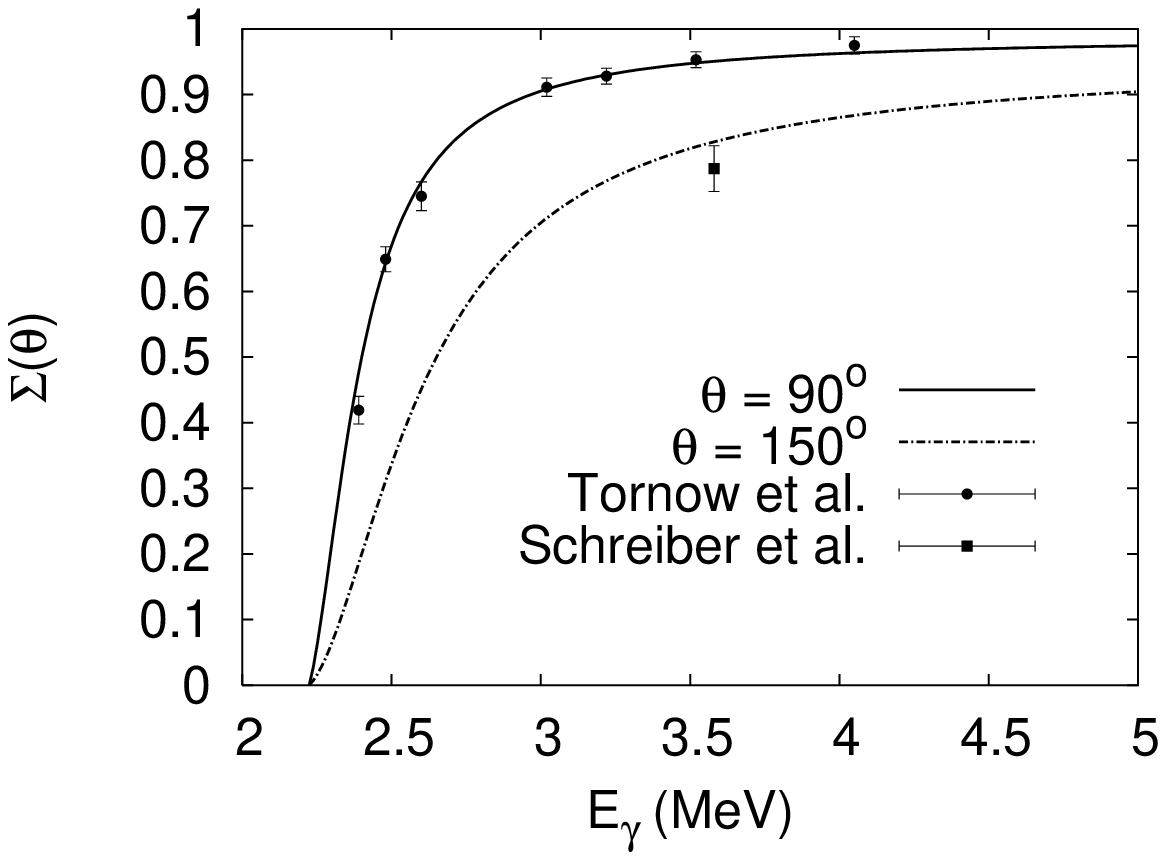, width=11cm}
\end{center}
\caption{
Photon analyzing power $\Sigma(\theta)$
for the $d + \gamma \rightarrow n + p$ process.
We plot our results at $\theta=90^\circ$ and $150^\circ$.
The experimental data are from Schreiber {\it et al.} ($\theta =150^\circ$)
\cite{setal-prc00} and Tornow {\it et al.} ($\theta = 90^\circ$)
\cite{tetal-plb03}.  }
\label{fig;anpower}
\end{figure}

First, we present our numerical results obtained by using the fitted
values of the parameters ``MCMC'' in Table~\ref{tab:parameters}, and
compare them with the experimental data
relevant to the BBN energies.
We plot in Fig.~\ref{fig;CSnpdg} the total cross section (in mb) of
$np \rightarrow d\gamma$ multiplied by the speed (in m/ns) of the
neutron in the laboratory frame as a function of the incident energy
of the neutron $E_n$.
We also plot the $M1$ and $E1$ contributions to the total capture cross
section in Fig.~\ref{fig;CSnpdg}, where the $M1$ contribution comes from
the amplitude of the initial $^1S_0$ state in Eq.~(\ref{eq;A1S0}) and
the $E1$ contribution from the amplitudes of the initial $P$-wave states
in Eqs.~(\ref{eq;A3P0},\ref{eq;A3P1},\ref{eq;A3P2}).
At very small energies, the $M1$ contribution overwhelms the $E1$
one. They become similar at around $E_n \sim 0.45$ MeV and after that,
the cross section is dominated by the $E1$ contribution.\footnote{
The contributions of $M1$ and $E1$ transitions to the total 
cross section of $np\to d\gamma$ and $d\gamma\to np$
are not disentangled in the measurements,
but the role of each amplitude can be studied from 
the measurement of $\Sigma(\theta)$.
}
The experimental data of $np\to d\gamma$ are compared with our result
in Fig.~\ref{fig;CSnpdg}.  Our results lie within the errors (5$\sim$6
\%) of all the data by Suzuki {\it et al.} and Nagai {\it et al.}.
%
In Fig.~\ref{fig;dgnp}, we plot our result of
the cross section of 
the $d\gamma\to np$ process and 
also separate the contributions from the $M1$ and $E1$ transition
amplitudes.
Recent measurement of the cross section at the BBN energies 
is reported by Hara {\it et al.}~\cite{hetal-prd03}.
An old datum by Moreh {\it et al.}~\cite{moreh-prc89} is also
included in the figure.
Our result 
agrees well with these experimental data within the error bars.
\footnote{
One may notice a departure 
(more than 1\%) 
of our estimation
from {\it the center values} of experimental data 
in Figs.~\ref{fig;CSnpdg} and \ref{fig;dgnp}.
This is because the curves plotted in the figures 
are mainly determined by the other data:
{\it e.g.},
the $np$ scattering data have a prime role to determine the 
energy dependence of the curves, whereas the accurate thermal $np$ capture
rates determine the normalization of them.  
However, as verified by the good $\chi^2$,
the curves are well within the experimental error bars. 
}
In Fig.~\ref{fig;anpower}, 
we plot our results of $\Sigma(\theta)$ 
at $\theta = 90^\circ$ and $150^\circ$
where experimental data are 
available~\cite{setal-prc00,tetal-plb03}.
We find good agreement between our predictions and 
the experimental data.
The error bars  
estimated for these quantities, discussed in detail below, 
are much improved compared to 
those of the experimental data.

Now, we compare our results with 
the predictions of the $np\to d\gamma$ cross 
sections at the BBN energies from various theories 
in Table~\ref{tab:theory-scaled}.
\begin{table}[tbp]
\begin{center}
\begin{tabular}{ c | c c | c c c }\hline
 & \multicolumn{5}{c}{Cross section (mb)} \\ \cline{2-6}
$E$ (MeV) &MCMC& Prev. Meth.
 &Rupak\cite{r-npa00} &Nakamura\cite{nakamura04} &Hale\cite{hale04}\\ \hline
$1.265\times 10^{-8}$&333.8(15) &333.7(15) &334.2(0) &335.0 &332.6(7)  \\
$5\times 10^{-4}$    &1.667(8) &1.666(8)  &1.668(0)  & 1.674   & 1.661(7)  \\ 
$1\times 10^{-3}$    &1.171(5) &1.171(5)  &1.172(0)  & 1.176   & 1.167(2)  \\ 
$5\times 10^{-3}$    &0.4979(23)&0.4976(21)&0.4982(0) & 0.4999  & 0.4953(11) \\ 
$1\times 10^{-2}$    &0.3322(15)&0.3319(14)&0.3324(0) & 0.3335  & 0.3298(9)  \\ 
$5\times 10^{-2}$    &0.1079(5)&0.1079(4) &0.1081(0) & 0.1084  & 0.1052(9)  \\ 
$0.100$              &0.0634(3)&0.0634(2)& 0.06352(5)& 0.06366 & 0.0605(10) \\ 
$0.500$              &0.0341(2)&0.0343(1)& 0.0341(2) & 0.03416 & 0.0338(8)  \\ 
$1.00$             &0.0349(3)&0.0352(1)& 0.0349(3) & 0.03495 & 0.0365(8)  \\ 
\hline 
\end{tabular}
\caption{
Theoretical predictions of the 
total cross section of the $n+p \rightarrow d + \gamma$ process
at the BBN energies. 
$E$ is the energy of two nucleons in the center of mass frame. 
See the text for details.
}
\label{tab:theory-scaled}
\end{center}
\end{table}
%
Our results, ``MCMC'' and ``Prev. Meth.'', in
Table~\ref{tab:theory-scaled} are calculated from the amplitudes in
dEFT up to NLO by a MCMC analysis and by using 
the values of the ``Prev. Meth.'' parameters
in Table~\ref{tab:parameters}, respectively.
We note that the 
values of the cross section 
in the column ``MCMC'' are
the most likely cross section values with 68.3\% central
confidence limits.
Thus we find that the error bars for the cross section are 
$\la$ 1\%\footnote{One should note that 
the error bars of the 68.3\% confidence
limits are different from (and larger than) those of the standard
deviation in multi-dimensional fits, whose typical 
error bars are $\la$ 0.3\%~\cite{a-thailand}.
}.
Values obtained
by Rupak~\cite{r-npa00} in Table~\ref{tab:theory-scaled} 
are from the pionless EFT (without dibaryon fields) 
up to N$^4$LO 
and those by Nakamura~\cite{nakamura04} 
are from the potential model calculation
using wave functions from the Argonne v18 potential
and meson exchange currents.
The results by Hale are obtained from 
an R-matrix analysis~\cite{hale04}.
There is good agreement between our two results,
``MCMC'' and ``Prev. Meth.'', 
up to 0.1 MeV,
whereas they show small differences ($\sim$0.6\%
and 0.9\%)
at $E =$ 0.5 and 1 MeV.
We also find good agreement 
between the ``MCMC'' analysis  
with that of the pionless EFT up to 
N$^4$LO by Rupak ($\la$ 0.2\%) 
and with that of the accurate potential model including the 
exchange current by Nakamura ($\la$ 0.5\%), 
while the results of the R-matrix theory at 
$E=$ 0.1 and 1 MeV significantly differ 
from the other estimations by $\sim$4.6\%.

Finally, we determine and present a
thermal capture rate and relative error
taking the recommended cross section from ``MCMC'' 
in Table~\ref{tab:theory-scaled}
for use in BBN computer codes:
\bea
f &=& N_A\langle \sigma v\rangle 
= 44216.0 [{\rm cm^3 \ s^{-1} \ g^{-1}}] (
1 
+ 3.75191 \, T_9 
+ 1.92934 \, T_9^2
\nonumber \\ &&
+ 0.746503 \, T_9^3 
+ 0.0197023 \, T_9^4 
+ 3.00491 \times 10^{-6} \, T_9^5) 
\nonumber\\ &&
/(1 
+ 5.4678 \, T_9 
+ 5.62395 \, T_9^2 
+ 0.489312 \, T_9^3 
+ 0.00747806 \, T_9^4)\, , 
\\
\delta f/f &=& 0.00449213(
1 
+ 3.08947 \, T_9 
+ 0.13277\, T_9^2 
+ 1.66472\, T_9^3) 
\nonumber \\ &&
/(
1 
+ 2.75245\, T_9 
+ 1.40958\, T_9^2 
+ 0.8791\, T_9^3)\, ,
\eea
where 
$T_9$ is a dimensionless 
temperature
defined by
$T_9= T/(10^9 K)$.
Using the thermal rates for the $np$-capture reaction
from ``MCMC dEFT'', Rupak, and Hale,
we show in Table~\ref{tab:lgtelab} how the light element
abundance predictions vary with the different $np$-capture cross sections.
We find little change for the mass fraction 
of $^4$He, $Y_p$ and the mole fractions $^3$He/H, 
a tiny ($\sim0.6$\%) change for D/H, and 
a small ($\sim2.9$\%) change for $^7$Li/H 
in the predicted light element abundances.
Differences between the light element predictions are not 
significant compared to the current estimated errors
in the BBN predictions.
In fact, the error budget in the BBN predictions is dominated by the
errors in other reactions such as
$d(p,\gamma)^3$He and $^3$He$(\alpha,\gamma)^7$Be 
(see, {\it e.g.}, Ref.~\cite{c-prd04} for more details).  
We have verified that the $np$-capture rate is not yet 
a major source of uncertainty in the light element abundance predictions.

\begin{table} 
\begin{center}
\begin{tabular}{l|c|c|c|c} \hline
$np$-capture rate & $Y_p$ & ${\rm D} / {\rm{H}} \times10^5$ & $\he3/{\rm H}\times10^6$ & $\li7/{\rm H}\times10^{10}$ \\ \hline
MCMC dEFT & 0.24852 & 2.5467 & 10.0921 & 4.4646 \\ \hline
Rupak~\cite{r-npa00} & 0.24853 & 2.5434 & 10.0920 & 4.4902 \\ \hline
Hale~\cite{jh-npa01} & 0.24849 & 2.5580 & 10.0864 & 4.3632 \\ \hline
\end{tabular}
\caption{The table shows how the light element abundance predictions 
vary with different $np$-capture cross sections.  
The nuclear rate
compilation from~\cite{c-prd04} was adopted, varying only the
$np$-capture rate.  In fact, these predictions follow the simple
abundance scalings (Eqs. (44-47) in Ref.~\cite{c-prd04}), 
using relative values
of the $np$-capture cross sections at the end of the D bottleneck
($E\sim0.07$ MeV). }
\label{tab:lgtelab}
\end{center}
\end{table}

\vskip 3mm \noindent
{\bf 6. Conclusions and discussion}

In this work, 
we calculated the total cross sections for
the $np \rightarrow d\gamma$ and $d\gamma \rightarrow np$
processes and the photon analyzing power in 
the $d\vec{\gamma}\to np$ process
at the energies relevant to BBN.
The pionless EFT that incorporates dibaryon fields 
was employed, and the transition amplitudes were calculated 
up to NLO. 
The values of the parameters 
(equivalently the LECs in effective Lagrangian)
and their uncertainties are evaluated
by a MCMC analysis ($\chi^2$-fit) using 
the relevant low energy experimental data.
Comparing our results with the experimental data and the 
previous theoretical estimations, 
we find good agreement within the estimated uncertainties ($\la$1\%)
except for the $np$-capture rate estimated 
by the R-matrix analysis 
at $E=$ 0.1 and 1 MeV where $E$ is the initial energy 
for two-nucleon in the CM frame.
These two values estimated by the R-matrix theory are 
considerably different from the other theoretical estimations 
by $\sim$ 4.6\%.
Therefore, it would be important to experimentally measure
the $np$-capture cross sections at these energies to resolve
this significant discrepancy. 

Now we are in the position to discuss 
the uncertainties 
of 
$np\to d\gamma$ cross section at BBN energies.
As discussed earlier,
the EFT calculations provide 
model-independent expression for the amplitudes
with a systematic perturbative scheme.
In the pionless EFT calculation up to N$^4$LO by Rupak,
various corrections in the higher order terms 
have been taken into account and it has been concluded
that the {\it theoretical} uncertainty up to the N$^4$LO
calculation in the $np\to d\gamma$ cross section 
is less than 1\%.
The effective range corrections are resummed in dEFT,
and so the convergence of dEFT is better than 
the pionless EFTs. 
Thus we expect that the contributions from
the higher terms, {\it i.e.}, 
the theoretical uncertainties in our calculation,
to be less than Rupak's estimation.   
With the overall good agreement between our MCMC analysis and the
Rupak and Nakamura calculations, the conclusion that the theoretical
uncertainties in the $np\rightarrow d\gamma$ cross sections is
$\la$1\% is well justified.  
The disagreement with the R-matrix analysis could 
be inferred from problems that R-matrix theory has in describing
non-resonant reactions.

Most of the recent theoretical calculations, as seen in
Table~\ref{tab:theory-scaled}, predict similar results with similar
uncertainties, and the accuracy of the calculations is better than the
$np$-capture experimental results at present.
This is the case mainly because 
there are a lot of accurate experimental data 
for the $np$ scattering and,
we could accurately constrain 
the four effective range parameters 
from them. 
We had only one additional LEC $l_1$ to fit from the low energy
$np\to d\gamma$ and $d\gamma\to np$ cross sections and the
photon analyzing power for the $d\vec{\gamma}\to np$ process.  
This situation, however, will change once we start studying 
other processes involving more than two nucleons  
for BBN, facing  
significant model dependence 
and lack of experimental data.
As discussed in Ref.~\cite{vk-npa05}, 
the EFT approaches would be useful in studying few-body
nuclear astrophysical processes 
because it provides simple model-independent expressions 
of the amplitudes with a finite number of LECs 
as well as a systematic expansion 
scheme.\footnote{Recently, 
the $nd\to \mbox{${}^3$H}\gamma$ process has been studied 
in the pionless dEFT~\cite{sb-npa05}.}
We expect that the combination of dEFT and the MCMC analysis 
can be a useful tool 
to estimate reliable uncertainties 
of few-body
nuclear reactions for BBN 
with the aid of relevant low energy experimental data. 

\vskip 3mm \noindent
{\bf Acknowledgments}

We are thankful to S.~X. Nakamura and G.~M. Hale for 
providing us with their numerical results.
SA thanks J.-W. Chen, H.~W. Hammer, L. Platter, 
Y. Nagai, A. Hosaka, K. Kubodera, 
D.-P. Min, M. Rho, and T. Kajino for discussions
and communications.
SA also thanks the Institute for Nuclear Theory 
at the University of Washington for its hospitality 
and partial support during the completion of this work.
SWH thanks B.~K. Jennings and TRIUMF for hospitality
during his sabbatical leave.
SA is supported
by Korean Research Foundation and The Korean Federation of 
Science and Technology Societies Grant funded by
Korean Government (MOEHRD, Basic Research Promotion Fund).
This work is supported in part
by the Natural Sciences and Engineering
Research Council of Canada and
by Faculty Research Fund, Sungkyunkwan University, 2004.

\vskip 3mm \noindent
{\bf Appendix A}

In this appendix we review the derivation of the 
renormalized dibaryon propagator from the low energy
$S$-wave $NN$ scattering~\cite{bs-npa01,ah-04}.

The LEC's $\sigma_{s,t}$ and $y_{s,t}$ 
in the effective Lagrangian, Eqs.~(\ref{eq;Ls}) and (\ref{eq;Lt}),
can be fixed from the effective range parameters of the $np$ scattering
in $^1 S_0$ and $^3 S_1$ states.
\begin{figure}
\begin{center}
\epsfig{file=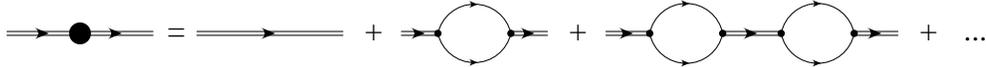,width=13cm}
\caption{
Diagrams for ``dressed'' dibaryon propagator
at leading order.
A double (single) line stands for a dibaryon (nucleon) field.}
\label{fig;dprop}
\end{center}
\end{figure}
Firstly, we derive ``dressed'' dibaryon propagators.
LO diagrams for the dressed dibaryon propagators 
in the $S$ wave channels are depicted in Fig.~\ref{fig;dprop}.
Since the insertion of the two-nucleon one-loop diagram 
does not alter the order of the diagram, the two-nucleon 
bubbles in the propagators should be summed up to 
infinite order. Thus the inverse dressed dibaryon
propagators for the spin singlet ($s$) ($^1S_0$) 
and triplet ($t$) ($^3S_1$) channels 
in the center-of-mass (CM) frame read
\bea
iD^{-1}_{s,t}(p) &=& i\sigma_{s,t}(E+\Delta_{s,t}) 
+ iy_{s,t}^2 \frac{m_N}{4\pi}(ip)
\nnb \\ &=&
i\frac{m_Ny_{s,t}^2}{4\pi}\left[
\frac{4\pi \sigma_{s,t}\Delta_{s,t}}{m_Ny_{s,t}^2}
+ \frac{4\pi \sigma_{s,t}E}{m_Ny_{s,t}}
+ ip \right]\, ,
\eea 
where we have calculated the two-nucleon one-loop diagram
using the dimensional regularization. 
$p$ is the magnitude of the nucleon momentum 
in the CM frame, and $E$ is the total energy $E\simeq p^2/m_N$.

\begin{figure}
\begin{center}
\epsfig{file=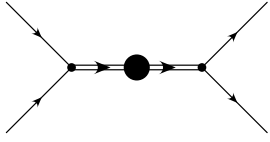,width=3.5cm}
\caption{
Diagram for the $NN$ scattering amplitudes.
The $dNN$ vertex is proportional to $y_{s,t}$
and the propagator of the dressed dibaryon field 
(a double line with a filled circle) is 
obtained from the diagram in Fig.~\ref{fig;dprop}.}
\label{fig;Amp}
\end{center}
\end{figure}

The $S$-wave $NN$ scattering amplitudes for both spin channels 
obtained from Fig.~\ref{fig;Amp}
read
\bea
i{\cal A}_{s,t} = (-iy_{s,t})(iD_{s,t}(p))(-iy_{s,t})
= \frac{4\pi}{m_N}\frac{i}{
-\frac{4\pi\sigma_{s,t}\Delta_{s,t}}{m_Ny_{s,t}^2}
-\frac{4\pi\sigma_{s,t}}{m_N^2y_{s,t}^2}p^2 -ip}\, ,
\label{eq;A}
\eea
and they are related to the $S$-matrix via
\bea
S_{s,t}-1 = e^{2i\delta_{s,t}} - 1
= \frac{2ip}{p\, {\rm cot}\delta_{s,t}-i p}
= i \left(\frac{p\, m_N}{2\pi}\right){\cal A}_{s,t}\, ,
\label{eq;A2}
\eea
where $\delta_{s,t}$ are the $S$-wave phase shifts.
The effective range expansion reads
\bea
p\, {\rm cot}\delta_s = -\frac{1}{a_0} + \frac{1}{2}r_0 p^2 + \cdots \, ,
\ \ \
p\, {\rm cot}\delta_t = -\gamma + \frac{1}{2}\rho_d (p^2+\gamma^2) 
+ \cdots \, ,
\label{eq;ERP}
\eea
for the $^1S_0$ and $^3S_1$ channel, respectively. 
Comparing the expressions of the amplitudes
in Eqs. (\ref{eq;A}) and (\ref{eq;A2}),
one has $\sigma_{s,t}=-1$ and 
\bea
&&
y_s = \frac{2}{m_N}\sqrt{\frac{2\pi}{r_0}}\, ,
\ \ \ 
D_s (p) = \frac{m_Nr_0}{2}\frac{1}{\frac{1}{a_0} 
 +ip -\frac12 r_0 p^2}\, , 
\\ &&
y_t = \frac{2}{m_N}\sqrt{\frac{2\pi}{\rho_d}}\, ,
\ \ \ 
D_s (p) = \frac{m_N\rho_d}{2}\frac{1}{\gamma +ip 
-\frac12 \rho_d(p^2+\gamma^2)}
= \frac{Z_d}{E+B} + \cdots \, ,
\eea
where $Z_d$ is the 
wavefunction normalization factor of the deuteron 
around deuteron binding energy $B$.
Ellipsis denotes corrections that are finite or vanish 
at $E=-B$. Thus one has
\bea
Z_d &=& \frac{\gamma\rho_d}{1-\gamma\rho_d}\, ,
\eea
which is the same as the asymptotic $S$-wave normalization
constant.

\vskip 3mm \noindent
{\bf Appendix B: Running a MCMC}

In this appendix we describe the steps in doing 
the MCMC analysis
for determining the $D=5$ parameters and cross sections.

1. Localizing the chain by minimizing the $\chi^2$.

The evolution of a Markov chain is governed by the $\chi^2$ values at
various points in parameter space.  
We adopt the standard definition:
\beq
\chi^2 = \sum_i \left( 
\frac{ \sigma_i({\rm thry}) - 
\sigma_i({\rm expt}) }{\delta\sigma_i({\rm expt}) } 
\right)^2,
\eeq
where $\sigma_i({\rm expt})$ and $\delta\sigma_i({\rm expt})$ are the
experimentally measured values ({\it e.g.} cross sections) 
and their total errors, respectively, 
and $\sigma_i({\rm thry})$ is the evaluated
theoretical value; the sum is over all data points.

Before we evaluate the $\chi^2$, we should pick some reasonable model
space.  Generally relying on previous works or positive definiteness,
one can place limits on the allowed parameter space.  Of course one
can expand this model space if a minimum is found at an edge of the
parameter space.  

We begin by selecting an initial set of parameters ($\vec{p}_0$) and
estimate step sizes ($\vec{\delta p}_0$) to be some fraction of the
size of each parameter space direction.  Drawing a set of random
numbers $\vec{z}$ with zero mean and unit variance ({\it e.g.} a
Gaussian-normal random number), one can determine a new set of
parameters via the relation:
\beq 
p = p_0 + \delta p_0z
\, .
\eeq
Evaluating the
$\chi^2(\vec{p})$ of this new set of parameters, and comparing to the
original $\chi^2(\vec{p}_0)$ we can determine if the new parameter
values better describe the data.  If the new $\chi^2$ is smaller we
accept the new parameter values ($\vec{p}_0 = \vec{p}$), otherwise
keeping the original parameter values.  We calculate new parameter
values and repeat.  The parameter values will gradually evolve to the
minimum possible $\chi^2$, by incrementally decreasing the step size
one can determine the best fit to some desired accuracy.

2. Checking for convergence to unique minimum.

By repeating this procedure with different starting parameter values
in our model space we can determine the uniqueness of this minimum.
This is particularly useful in testing the boundaries of the chosen
model space.  If a minimum is found on a boundary the parameter space
must be enlarged.  If any 2 parameters are completely correlated, a
unique minimum will not be found 
({\it e.g.} $l_1$ and $l_1^{\prime}$)\footnote{
$l_1^{\prime}$ is a LEC associated 
with a vector-dibaryon-nucleon-nucleon vertex.
The LECs $l_1$ and $l_l^{\prime}$
are almost redundant in the $np$-capture cross section~\cite{ah-04}.
} 
and
one must reconsider the allowed parameter space.  We will assume from
now on that there is only one minimum in our model space and that no
two parameters are completely correlated.

3. Determining an appropriate step-size.

To make the MCMC as efficient as possible, one needs to determine an
appropriate step size, $\vec{\delta p}$.  A simple method of
estimating this is by varying individual parameter values away from
the minimum, until the difference
$\chi^2-\chi^2_{\rm min}$ is unity.  This choice makes for a
good first estimate, and will generally be smaller than or equal to
the true parameter errors.

4. Running the chain(s).  Enforced constraints.

There are many ways to initiate a MCMC.  Some choose to pick an
initial point randomly in the model space, while others choose to
initiate the chain at the minimum.  The latter method avoids the
dependence on the prior probability distribution the initial point is
generated from and thus reducing the overall convergence length of the
chain.

Once the initial point of the chain is chosen, we follow a procedure
quite similar to that used in the minimization algorithm.  One
generates a new set of parameters just as in the minimization
algorithm.  There is then a set of criteria for accepting or rejecting
the new parameter set:
\begin{enumerate}
\item
if $\Delta\chi^2 = \chi^2(\vec{p}) - \chi^2(\vec{p}_0) < 0$ we accept the new parameter values.

\item
otherwise there is a finite probability, $P=\exp{(-\Delta\chi^2/2)}$ for
accepting the point.

\end{enumerate}

For a reasonable step size choice, this allows for the efficient
exploration of the ``tail'' of the parameter likelihood distribution.
Whether or not one accepts the new point, 
a new set of parameters is drawn 
and this step is repeated until the chain has met its
convergence criteria or some maximum length.

5. Checking convergence to a ``full'' sample.

A relatively simple method to check the convergence of a single chain
is to calculate the $1^{st}$ and $2^{nd}$ order moments of the chain of a
specific length $N$.  We thus calculate the means, variances
and covariances of the $D$ parameters:
\bea
\vec{\mu}(N) &=& \frac{1}{N} \sum_i^N \vec{p}_i \, , 
\\
{\mathcal C}(N) &=& \frac{1}{N-D} \sum_i^N [\vec{p}_i-\vec{\mu}(N)]\otimes[\vec{p}_i-\vec{\mu}(N)],
\eea
where the variances are the diagonal components of the covariance
matrix. As the MCMC converges, these moments of the underlying
likelihood distribution will plateau and the fractional difference
between the $N^{th}$ and $(N+1)^{th}$ moments (or functions of the
moments, {\it e.g.} 
det(${\mathcal C}$)) should fall like $\sim1/N$.  Thus, we
choose $N$ in such a way to reach a certain desired fractional
uncertainty.

One can also compare multiple chains and their moments.  One can then
compare the variance of a single chain to the variance of the chain
means, adopting the convergence criteria from~\cite{convergence}.
However, since we are starting our chains at the minimum, a single
chain is all that is needed once the chain has grown longer than the
intrinsic correlation length of the chain (typically $\sim$ few 100's
steps) and the convergence criteria for the single chain is all that
is needed.

To meet convergence criteria, chains with length $N\sim10^6$ are
required.  The parameter step size is adjusted so that the acceptance
is between 25 and 50\%.  We find that a 40\% acceptance rate is most
efficient.

\begin{table} 
\begin{center}
\begin{tabular}{l|l|l|l} \hline
best fit values/step-sizes (fm) & minimum $\chi^2$ & $<\!\!\chi^2\!\!>\!\pm\,\delta\chi^2$ & \# data \\ \hline
$a_0\pm\delta a_0 = -23.745\pm0.0070$  & $\chi^2_{\rm cap,min} = 4.63$ & $5.69\pm1.48$ & $N_{\rm cap} = 6$ \\ \hline
$r_0\pm\delta r_0 = 2.732\pm0.0095$    & $\chi^2_{\rm dis,min} = 4.55$ & $4.55\pm0.48$ & $N_{\rm dis} = 9$ \\ \hline
$\rho_d\pm\delta \rho_d = 1.740\pm0.0015$ & $\chi^2_{\rm sct,min} = 2282.83$ & $2285.99\pm2.63$ & $N_{\rm sct} = 2124$ \\ \hline
$l_1\pm\delta l_1 = 0.894\pm0.0012$    & $\chi^2_{\rm anp,min} = 10.99$ & $10.99\pm0.25$ & $N_{\rm anp} = 8$ \\ \hline\hline
    &  $\chi^2_{\rm tot,min} = 2303.00$ & $2308.22\pm3.22$ & $N_{\rm tot}=2147$ \\ \hline
\end{tabular}
\caption{This table shows the parameter values at the minimum total $\chi^2$.  
The step-sizes should not be confused with error bars. Also evaluated
about this point are the step-sizes for each parameter.  The second
and third columns show the contribution to the minimum and average
$\chi^2$ values from each of the types of data (``cap" = capture,
``dis" = dissociation, ``sct" = scattering, ``anp" = analyzing power and
``tot" = total $\chi^2$'s, respectively) and the fourth column the number
of data points evaluated.}
\label{tab:minchi2}
\end{center}
\end{table}

\end{document}